# Android Malware Detection: an Eigenspace Analysis Approach

Suleiman Y. Yerima, Sakir Sezer
Centre for Secure Information Technologies (CSIT)
Queen's University Belfast
Belfast, Northern Ireland
{s.yerima, s.sezer}@qub.ac.uk

Igor Muttik
Sr. Principal Research Architect
McAfee Labs (Part of Intel Security)
Aylesbury, Buckinghamshire, United Kingdom
mig@mcafee.com

***Abstract*—The battle to mitigate Android malware has become more critical with the emergence of new strains incorporating increasingly sophisticated evasion techniques, in turn necessitating more advanced detection capabilities. Hence, in this paper we propose and evaluate a machine learning based approach based on eigenspace analysis for Android malware detection using features derived from static analysis characterization of Android applications. Empirical evaluation with a dataset of real malware and benign samples show that detection rate of over 96% with a very low false positive rate is achievable using the proposed method.**

## I. INTRODUCTION

Mobile malware has become an issue of increasing concern, particularly on the Android platform which has been targeted relentlessly in recent years. In contrast to other mobile platforms, Android allows installation of apps from third-party markets and other unverified sources making it more susceptible to malware. The exponential growth in Android devices and the buoyant and largely unregulated app market produced a sharp rise in malware targeting the platform. In 2013, an estimated 1000 new Android malware samples were discovered every day; in 2014 the number has risen to 2000 malware samples seen daily [1].

Since their debut in mid-2010, Android malware have evolved beyond simple information stealing programs to more sophisticated malicious software employing various obfuscation techniques and detection avoidance capabilities. For example, sophisticated Android malware families are employing payload encryption, obfuscated command and control channels, runtime dynamic loading of malicious payload, etc. which makes detection and analysis more challenging.

Another major problem is the difficulty of spotting malware in the wild, given that many app sources exist and thousands of apps are uploaded every day. Traditional scanners that rely on malware signatures are not effective at detecting new malware for which they have no existing signature in their database. Hence, new techniques to enable timely and proactive discovery of unknown Android malware are definitely required.

Hence, in this paper, we propose and evaluate an Android malware detection scheme based on the face recognition technique known as eigenfaces [2]. The *eigenfaces* method measures the extent of similarities and differences amongst a set of faces in a database. It is based on the premise that every face is a linear combination of the basic set of faces called eigenfaces. The eigenfaces are projections of real faces into an *eigenspace* which is defined by the vectors that spans across significant variations amongst them. In this paper, these basic principles are applied to develop an eigenspace model based on static analysis characterization of applications which is then used to determine whether or not a given new application is potentially malicious.

In the next section related work in Android malware detection will be discussed; next, the approach proposed in this paper will be detailed, then the experimental studies undertaken to evaluate the effectiveness of the scheme will be presented, followed by analysis of the results, conclusions and future work.

## II. RELATED WORK

In this section, an overview of related prior works in Android malware detection that are based on static analysis and machine learning is presented. Static analysis is a method of scrutinizing an application without execution, in order to uncover potentially malicious behaviour. It underpins the approach proposed in this paper. Several works have utilized static analysis for Android malware detection. Some of these proposals are based on heuristics or manually crafted rules for example [3] and [4]. Others have been designed to detect vulnerabilities e.g. Comdroid [5] and DroidChecker [6].

Recently, several other efforts have combined static analysis with machine learning for Android malware detection. For example, DroidMat is proposed in [7] as a static feature-based system that applies K-means algorithm for enhanced modelling and the kNN algorithm to classify applications as benign or malicious. The DroidMat system used features based on requested permissions, Intents, and API call tracing for each app component. The authors compare their system to the Androguard malware analysis tool using 1738 sample applications.

In [8] a Bayesian classification model is developed using static analysis based on 58 features derived from API calls,

intents, and system commands to classify Android applications as ‘benign’ or ‘suspicious’. This model was further analysed in [9] to study the impact of using features derived from permissions alongside API calls, intents and system commands for Bayesian-based detection of Android malware. Peng et. al. [10] also proposed and evaluated probabilistic generative models based on Android permissions features; while Sanz et. al. [11] trained and compared several machine learning algorithms employing Android permissions features in order to detect malware. X. Liu and J. Liu [12] proposed a two layer system based on permission features that classify applications in three stages, with each stage utilizing the J48 Decision Tree algorithm to determine whether an application is suspicious or benign. Sahs and Khan [13] utilized call flow graphs to train a one-class SVM machine learning algorithm for detection of Android malware, while Arp et. al. [14] proposed DREBIN which also employed SVM and performs a broad static analysis to gather as many features of an application as possible for device-based detection of suspicious apps.

Other recent efforts include: Sharma and Dash [15] where models based on API calls and requested permissions were built and investigated using Naïve Bayes and K-Nearest Neighbour classifiers. Yerima et. al. [16], investigated parallel classifiers that utilized static features based on permissions, system commands and API calls. The paper compared various parallel classification combination schemes aimed at improving detection accuracy over single classification algorithms. Unlike [7]-[16] however, Kang et. al. [17] did not use features from decompiled or disassembled code but rather utilized Dalvik bytecode frequency analysis extracted from 1300 malware samples and classified the malware into their 26 respective families using the Random Forest machine learning algorithm.

Different from all the aforementioned existing works, this paper proposes and evaluates an eigenspace approach for Android malware detection. Earlier work employing related approaches for malware detection were investigated for Windows-based metamorphic worms by Saleh [18] and Deshpande [19]. A different approach is developed in this paper for Android malware detection using the same basic principles.

## III. EIGENSPACE APPROACH FOR ANDROID MALWARE DETECTION

A set of input features are required to characterize the applications when using the eigenspace approach. The features chosen to characterize the applications are motivated by the manner in which malware typically misuse certain API calls, permissions, and intents for malicious activities, and how they attempt to illegitimately run system commands for privilege escalation. The features are extracted from static analysis of the APKs in order to characterize each application according to their usage of these features.

### A. *Applications characterization*

The features extracted from the applications for characterization and input into the eigenspace analysis based system consists of: API related features, permissions related features, intents, and commands related features. Previous works such as [7]-[12] and [14]-[16] have found these type of static features to be effective for training machine learning models for classification. Hence, we are similarly motivated to utilize features derived from these categories to develop the eigenspace based detection models proposed in this paper.

The API related features are obtained by mining the disassembled Dalvik executable (dex) file. These include API calls for accessing subscriber identity, device identity, executing external commands, intercepting broadcast notifications, encryption, etc. Indeed, most mobile malware attempt to steal sensitive information or send premium SMS messages by taking advantage of standard platform APIs [20]. It is therefore conceivable to include features related to such API calls in our feature set.

The permissions related features are keywords that map onto standard Android permissions which an app requests by declaring them in the manifest file. As observed previously, (e.g. [9], [11], and [12]) certain permissions are requested more frequently by malware than others. For example, ‘risky’ permissions that request ability to read contacts, read SMS messages, send SMS messages, install packages, delete packages, access location information, etc. are commonly found to be declared in manifest files of packages containing malware.

Intents are used for intra-process and inter-process communication on Android. They are passive data structures exchanged as asynchronous messages allowing information about events to be shared between different applications and different components of applications. For example, malware commonly listen for the BOOT_COMPLETED intent in order to trigger malicious activity immediately on booting a device.

Commands related features are keywords that detect the presence of system commands like ‘chown’, ‘chmod’, ‘mount’ etc. or certain parameters which may be used with these commands. In many malicious APKs, these commands can be found embedded in hidden files within an APK and invoked to enable privilege escalation, launch concealed scripts, remove traces of malicious activities, install additional malicious components, etc. The features used for characterizing applications for the eigenspace scheme developed in this paper can be found in the appendix.

### B. *Model development*

The goal is to *apply the eigenspace approach* to recognize whether a given unknown Android application is potentially malicious or not based on how close it is *determined to be similar to a known application* in the dataset and then assign it to the same class as the known application. A set of malware and benign applications are used *to create an eigenspace* for detecting new/unknown malware.

The training sets were constructed as follows:

1. Create a training set $S$ consisting of $M$ malware and $B$ benign applications. Denote $S = \{S_1, S_2, ..., S_K\}$ where $K = M + B$
2. Let $N$ be the number of features characterizing each application in set $S$. Each application is represented by its features in an $N \times 1$ column vector $\boldsymbol{V}$ where $\boldsymbol{V}^{\mathbf{T}} = [f_1, f_2 \ldots f_N]$ and $f_i \in \{0,1\}$ for $i = 1 \ldots N$

3. Arrange all the individual column vectors into an $N \times K$ matrix $\boldsymbol{\Phi} = [V_1, V_2, \ldots, V_K]$
4. Compute $\boldsymbol{\varepsilon}$, the column vector representing the mean features of the set of the training applications represented in $\boldsymbol{\Phi}$ as follows:

$$\varepsilon = \frac{1}{K}\sum_{n=1}^{K} V_n \quad (1)$$

5. Obtain the matrix $\boldsymbol{A} = [V'_1, V'_2, ..., V'_K]$ where $V'_i = V_i - \varepsilon$, for $i = 1 \ldots k$
6. Next, find the eigenvectors $\boldsymbol{\gamma}$ of the covariance matrix $\boldsymbol{C}$, where $\boldsymbol{C} = \boldsymbol{A}\boldsymbol{A}^T$. Note that $\boldsymbol{C}$ is an $N \times N$ matrix, and since $N \ll K$ given the number of applications to be incorporated into the training set is larger than the number of features, the eigenvectors are computed directly from $\boldsymbol{C}$.
7. Suppose $\gamma_i$ is a eigenvector of $\boldsymbol{C}$ then

$$\boldsymbol{C}\,\gamma_i = \boldsymbol{\lambda}_i \gamma_i \quad (2)$$

where $\boldsymbol{\lambda}_i$ is the corresponding eigenvalue.

8. Thus the set of eigenvectors $\boldsymbol{\gamma} = \{ \gamma_1, \gamma_2, \ldots, \gamma_{N'} \}$ are computed and sorted in descending order of magnitude of their corresponding eigenvalues with those having higher values being more important in describing the applications. Hence, a number of $\boldsymbol{N'}$ eigenvectors are chosen to describe the *eigenspace* (i.e. the linear space spanned by the selected eigenvectors).
9. Each sample in the training set $S$ is projected into the eigenspace defined by these eigenvectors by representing each one as a linear combination of eigenvectors and weights.

$$V'_i = \sum_{j=1}^{N'} w_j \gamma_j, \quad \text{where } N' \le N \quad (3)$$

10. The weights of each application in the training set can be calculated from

$$w_j = \gamma_j^T V'_i, \quad j = 1, 2, ..., N' \quad (4)$$

11. The weights of the application can be combined into a vector $\boldsymbol{W}$, where $\boldsymbol{W}^T = [w_1, w_2, \ldots, w_{N'}]$
12. Let $\boldsymbol{D} = [W_1, W_2, \ldots, W_K]$
13. $\boldsymbol{D}$ can be considered a model trained from the $K$ samples in $S$ which consists of applications from both malware and benign classes.

### C. *Classifying a new application:*

In order to classify a new application, its defining features $(f_1, f_2 \ldots f_N)$ are first extracted and steps 4 and 5 above are used to subtract the mean (training set) application. Thus we obtain $\boldsymbol{V'}_{new} = \boldsymbol{V}_{new} - \boldsymbol{\varepsilon}$ where $\boldsymbol{V}_{new}$ is the column vector of its defining features.

Next, we project $\boldsymbol{V'}_{new}$ into the eigenspace defined by the set of eigenvectors $\gamma_1, \gamma_2, \ldots, \gamma_{N'}$ to obtain its weight vector:

$$\boldsymbol{W}_{\boldsymbol{new}}^T = [w_1, w_2 \ldots w_{N'}] \quad \text{where} \quad w_j = \gamma_j^T V'_{new}, \quad j = 1, 2, ..., N'$$

Hence, to determine how much the new application is close to an application in the training set with weight vector $W_i$, the Euclidean distance between the weight vectors are calculated to yield a score given by:

$$\text{Score} = \|W - W_i\| = \sqrt{\left((w_1-\omega_1)^2 + (w_2-\omega_2)^2 + \ldots + (w_{N'}-\omega_{N'})^2\right)} \quad (5)$$

By scoring the new application against each application in the training set $S$ represented by their respective weight vectors in $\boldsymbol{D}$, we can predict the class of the application by:

$$P = \arg\min_i \|W - W_i\|, \quad i = 1 \ldots K \quad (6)$$

If P belongs in the (labelled) malware class then the new application is predicted to be malicious otherwise it is predicted to be benign.

## IV. Methodology and Experiments

This section describes the methodology of the experiments that were conducted to evaluate the Android malware detection approach proposed in this paper.

### A. *Dataset pre-processing*

The experiments were undertaken using 6,860 applications; of these, 2,925 were malware while the remaining 3,935 were benign applications. The samples were provided by McAfee (part of Intel Security). In order to extract the defining features from the applications, a bespoke Android package analysis tool was developed using Java and python. The tool enables automated reverse engineering of Android applications to allow for the construction of feature vectors which are subsequently arranged into a matrix of column vectors characterizing each application as explained in the previous section.

Initially, 175 features based on API calls, permissions, intents and commands related keywords were extracted for each of the applications. The features were ranked in order of relevance using the G*ain Ratio* criterion in WEKA [21]. Subsequently, 100 top ranked features were selected for training the eigenspace model. The 100 features are given in the appendix in their ranking order. Thus, according to our model, we have N = 100. This pre-processing stage resulted in a $100 \times 6860$ matrix of feature vectors that were further processed using various python scripts and MATLAB in order to build the eigenspace model as described in section III.

### B. *Model training phase*

The experiments were conducted using a 5-fold cross validation approach with 80% of the dataset used for training and 20% for testing. Thus, each time, a different testing set was used against a complementary trained eigenspace model consisting of 80% of the dataset and the results were averaged.

A training set consisted of feature vectors from 2340 malware and 3151 benign applications, giving an input matrix $\boldsymbol{\Phi}$ of dimension 100 × 5491 for the model. $\boldsymbol{\varepsilon}$ was calculated and subtracted to obtain matrix $\boldsymbol{A}$ of the same dimension. The set of eigenvectors and eigenvalues were computed as described in section III, and then the top 70 eigenvectors (out of 100) covering 95% variance were selected to construct the eigenspace. Thus, $N'$=70 with the eigenspace dimension reduced to 70 × 100 from the initial 100 × 100 matrix. Next, the weight vectors $\boldsymbol{W}$ are calculated by projecting each training feature vector (from $\boldsymbol{A}$) into the eigenspace defined by the $\boldsymbol{N'}$ eigenvectors. Finally, a matrix $\boldsymbol{D}$ is produced from the operation to yield a trained model against which the testing set is evaluated.

### *C. Testing and classification phase*

The test set for each of the 5 folds consisted of 585 malware samples and 787 benign samples (20% of the dataset). For each test set, the weight vectors $\boldsymbol{W}$ were computed against the corresponding training set's eigenvectors. During the testing phase, each test application was measured against each app in the training set using the Euclidean distances (of their projections into eigenspace represented by the weight vectors $\boldsymbol{W}$ ) to derive a score for each app in the training set. The class of the test app is predicted by assigning it to the class of the app that returns the minimum score in the training set. The app that returns the minimum score is considered to be the one that is (functionally) *most similar* to the test app from amongst the training set. In our experiments performed on a 64 bit Windows 7 system with an Intel Xeon 2.27 GHz CPU and 16GB RAM, the classification took between 2.1 to 2.2 seconds for each app in the testing set.

Since there are two classes to be predicted, the matrix $\boldsymbol{D}$ of the trained model, was arranged to have the weight vectors of the malware samples preceding those of the benign samples. Thus $W_1, W_2, \ldots, W_{2340}$ were malware vectors while $W_{2341}, W_{2342}, \ldots, W_{5491}$ represented benign weight vectors. Hence, 2340 was the decision threshold for determining whether a given test application was benign or malware. The test apps that returned a minimum score corresponding to an app labelled equal to or below 2340 were predicted to be malware. While those that returned a minimum score corresponding to an app labelled above 2340 were predicted to be benign. Hence, the classification criteria for a given new application $S_{new}$ are stated as follows:

If P=arg $\min_i \|W-W_i\| \leq 2340$ Then $S_{new}$ is Suspicious

If P=arg $\min_i \|W-W_i\| > 2340$ Then $S_{new}$ is Benign

Measurements of prediction accuracy were taken during each run of the experiment and averaged to obtain the overall results. The results are presented and discussed in the next section.

## V. Results and Discussions

In this section, the results of the experiments undertaken to evaluate the model are presented. Figure 1 depicts a mapping of test samples to training samples during one run of the experiments. This gives a visualization of the prediction accuracy at a glance. The x-axis is the test sample number while the y-axis is the training sample number. The dots represent an intersection of test and training sample pair that are closest in the eigenspace. The lines demarcate the labelled malicious samples from the benign ones; hence, the quadrants with the dense points correspond to correctly predicted classes while the two quadrants with the sparse fewer points denote the incorrectly predicted classes (i.e. false positives and false negatives). This illustrates that the eigenspace approach was quite effective at guessing the class of the test samples with only very few false positives and false negatives. Also, since only the weights are used, the determination of the class is quite fast during the testing phase, i.e. 2.1-2.2 seconds as mentioned earlier.

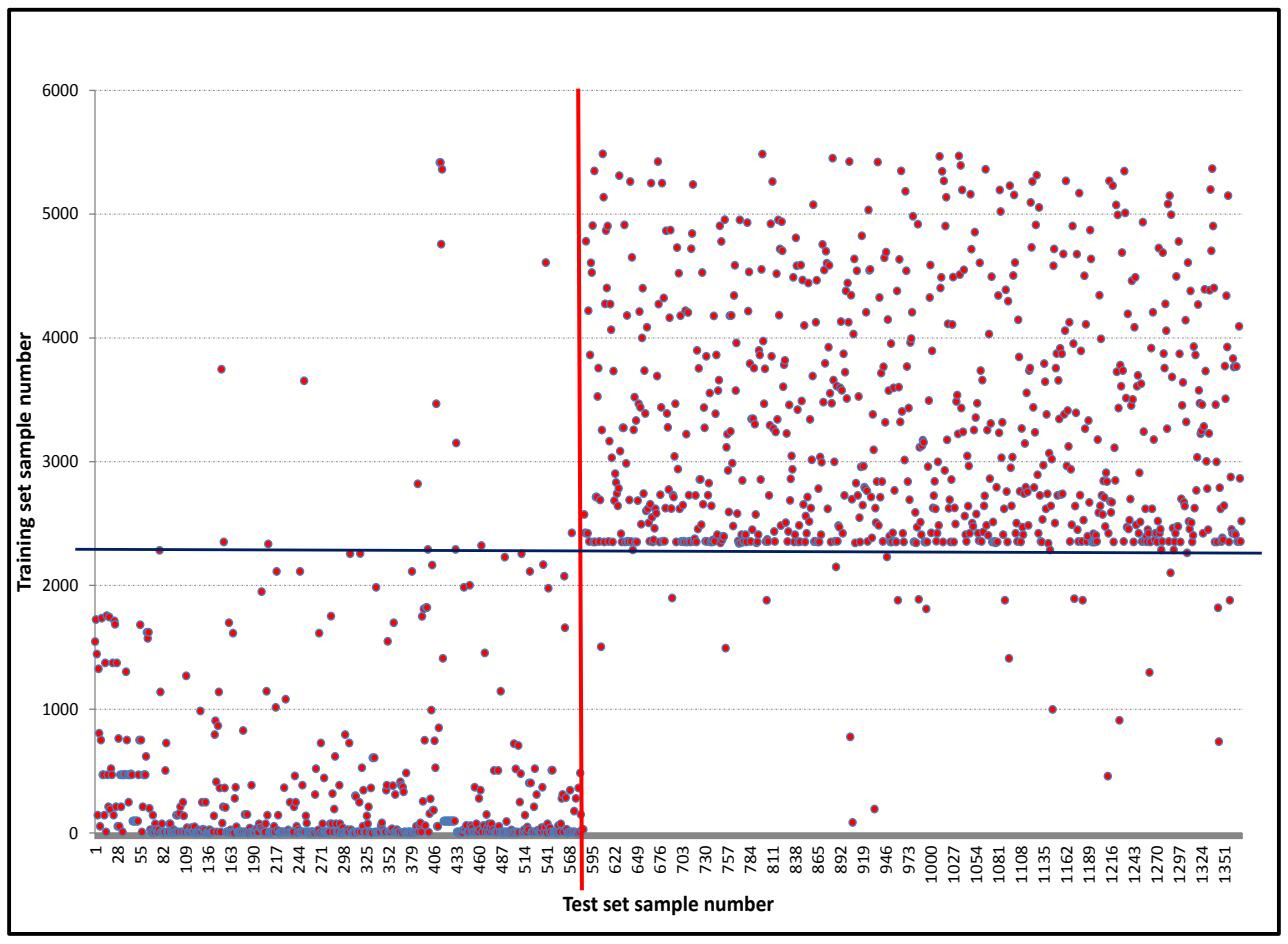

Fig. 1. Depiction of classification accuracy and error at a glance: Mapping test samples to training samples via the eigenspace method. Testing samples class (suspicious/benign) threshold is 585. Training samples class (suspicious/benign) threshold is 2340.

Table I presents the average accuracy measurements obtained in the evaluation of the eigenspace approach. Here, the TPR (True positive ratio) i.e. ratio of correctly classified malware to the total malware apps; TNR (True negative ratio) i.e. ratio of correctly classified benign apps to the total benign apps; FNR (False positive ratio) i.e. incorrectly classified malware; FPR (false positive ratio) i.e. incorrectly classified benign apps; overall accuracy (ACC) and error (ERR) are shown.

The table shows that **96.3%** of the malware apps were correctly classified i.e. only **3.7%** of the malware were false negatives. Also, **96.5%** of the benign apps were correctly classified by the eigenspace approach, with only **3.6%** recorded as false positives.

Table I. Performance evaluation of the eigenspace space approach vs. other classification methods

| | TPR | FPR | TNR | FNR | ACC | ERR |
|---|---|---|---|---|---|---|
| **Eigenspace** | **0.963** | **0.035** | **0.965** | **0.037** | **0.964** | **0.036** |
| **NB** | 0.772 | 0.087 | 0.913 | 0.228 | 0.843 | 0.158 |
| **DT** | 0.945 | 0.040 | 0.960 | 0.055 | 0.953 | 0.048 |
| **SL** | 0.901 | 0.049 | 0.951 | 0.099 | 0.926 | 0.074 |
| **PART** | 0.953 | 0.041 | 0.959 | 0.047 | 0.956 | 0.044 |

These results demonstrates that the proposed eigenspace approach is very effective for Android malware detection and is also able to classify the apps with improved classification accuracy compared to other popular machine learning techniques as seen from Table I. Using the same dataset and 5-fold cross validation approach, results from the Naïve Bayes (NB), Decision Tree (DT), Simple Logistic (SL) and PART learning algorithms were obtained for comparison to the eigenspace approach. Figures 2, 3 and 4 present a graphical depiction of the comparative results in the table.

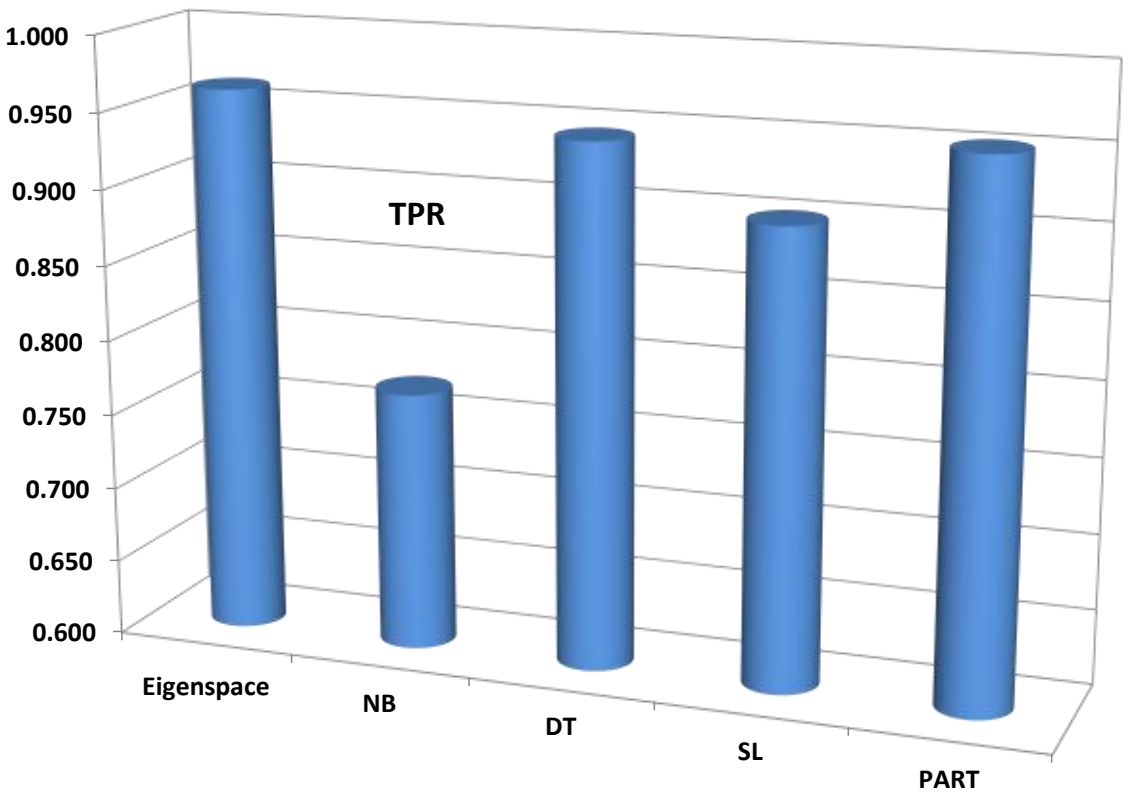

Fig. 2. TPR comparison of eigenspace approach with NB, SL, PART and DT machine learning algorithms.

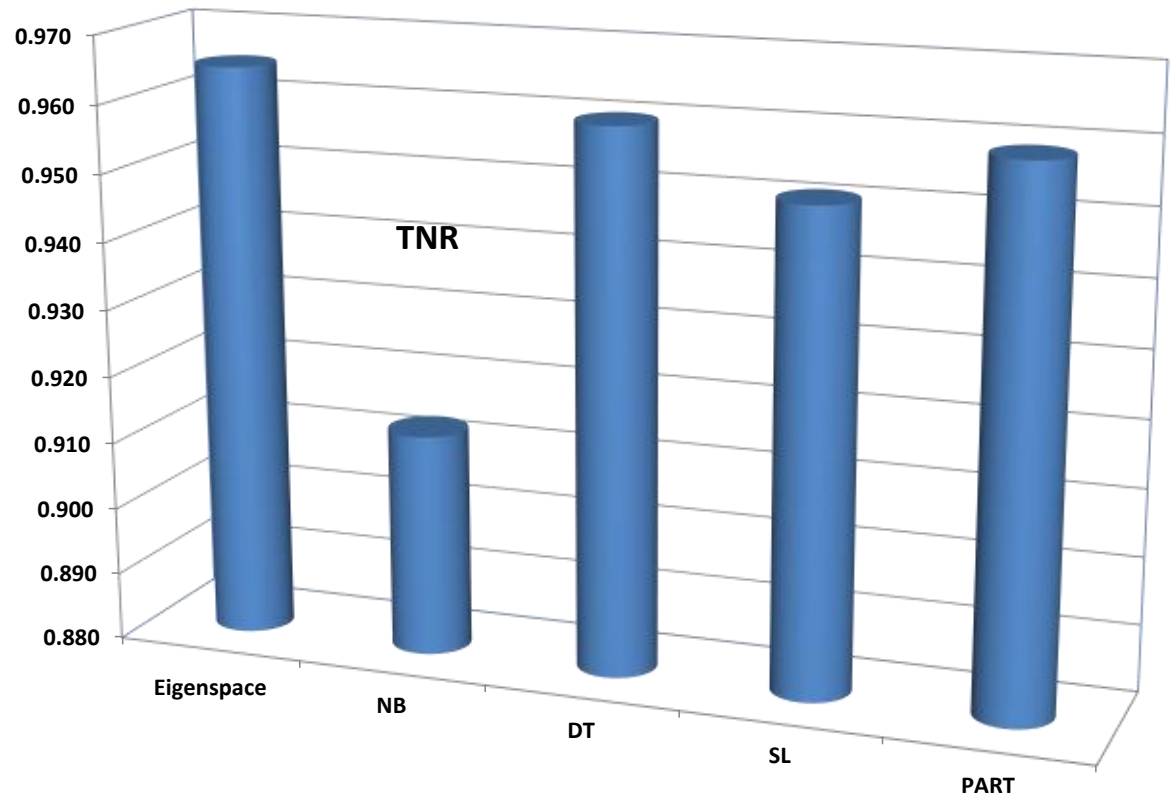

Fig. 3. TNR comparison of eigenspace approach with NB, SL, PART and DT machine learning algorithms.

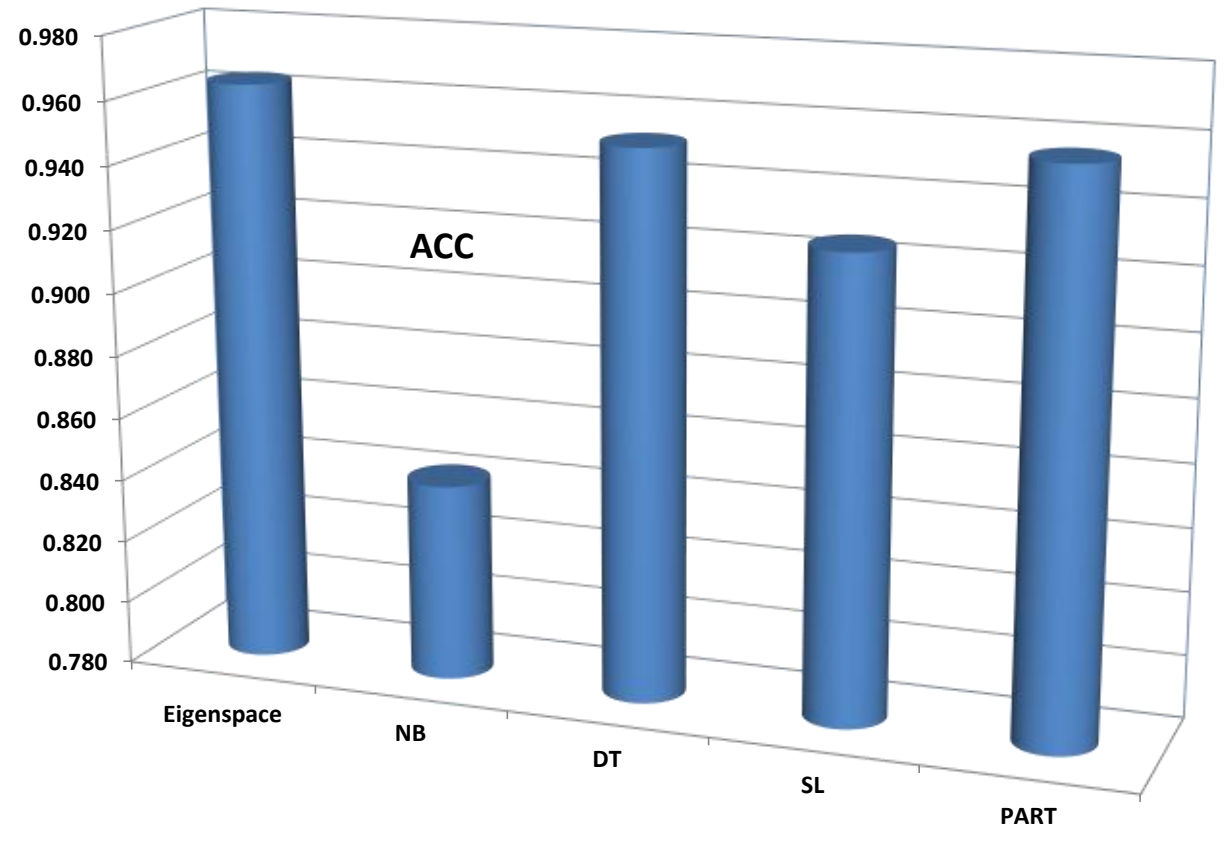

Fig. 4. ACC comparison of eigenspace approach with NB, SL, PART and DT machine learning algorithms.

From figure 2, it can be seen that the eigenspace approach was able to predict more malware samples accurately than the other four learning algorithms. Figure 3 also showed that the eigenspace approach was able to guess more of the benign samples correctly than the other four algorithms. The overall accuracy depicted in Figure 4 shows that despite how well the DT and PART algorithms work for the given features that have been used, the eigenspace method still surpassed them in overall prediction accuracy. Thus, we can also conclude that the particular set of static features used are well suited to building eigenspace models that can perform very well in predicting and classifying Android apps.

### *A. Discussions*

Decision Trees classification algorithms are known to perform very well on many datasets. For the experiments in this paper we used the WEKA implementations of the J48 decision tree algorithm. PART is a rule based approach that uses partial decision trees for rule induction; hence it is not surprising that that PART also shows a good accuracy performance like the DT. The eigenspace approach does better than either one of them, and in practice can be quite computationally cost effective in classification of new samples, especially when a streamlined training set is used with pre-computed Euclidean distances utilized for the 'minimum score' calculation.

Since the eigenspace approach attempts to determine the class by finding the closest or most similar app in the training dataset, it could potentially enable a more straightforward determination of the family a new app might belong to or an indication of what functionalities might be present, than the other machine learning methods. For instance, referring back to Figure 1, assuming that the families of the training samples on the y-axis and their functionalities are known, the mapping of the test sample to a training sample could be used to approximate the malware family or estimate possible functionalities (depending on the magnitude of the minimum score). In the same vein, another potential advantage compared to the other machine learning techniques is the possibility to infer significant differences in functionality of newly detected malware from existing families. Furthermore, eigenspace models can readily be improved over time by a self-improving mechanism that can automatically eliminate false positive clusters from the training set or include new true negative samples into the training set (to further enrich the eigenspace). This possibility to dynamically control the FPR and TNR in this manner is another advantage of the eigenspace approach over other machine learning methods. The ease which the model can be incrementally trained/re-trained to improve accuracy by including novel training samples makes it an attractive machine learning based classification technique for detecting new Android malware in practice.

## VI. CONCLUSIONS

This paper presented an effective eigenspace analysis approach to Android malware detection. The proposed approach is investigated using 2,925 real malware samples and 3,935 clean samples employing standard cross-validation method. The eigenspace approach is based on the eigenfaces technique which has its origins in face recognition applications. The results obtained from extensive empirical evaluations show that it is a promising scheme for Android malware detection with 96.4% accuracy and only 3.6% false positives observed. We have also found that compared to several popular machine learning techniques the eigenspace approach performs quite well and at the same time can enable better usability in practical systems. Moreover, it is easily applicable in many scenarios where inference of other additional knowledge such as related malware families may be useful.

The model proposed in this paper can be improved further. Hence, future work will investigate means of reducing detection errors, such as applying more effective filters at the feature extraction phase to improve the application characterization, or deriving and experimenting with different sets of features which could be more discriminative for classification. Another aspect for further investigation is streamlining and optimizing the eigenspace whilst still retaining high accuracy performance.

## ACKNOWLEDGMENT

The authors grateful to McAfee (part of Intel Security) for providing the sample applications that were used in this work.

## APPENDIX

TABLE II. 100 TOP GAIN RATIO RANKED FEATURES USED FOR THE EIGENSPACE BASED ANDROID MALWARE DETECTION MODEL.

| Feature | Type | Feature | Type |
|---|---|---|---|
| SEND SMS | P | BROADCAST SMS | P |
| createSubprocess | API | KILL BACKGROUND PROCESSES | P |
| remount | CR | READ SYNC STATS | P |
| /system/bin/sh | CR | CAMERA | P |
| chown | CR | res | CR |
| RECEIVE SMS | P | KeySpec | API |
| /system/app | CR | DELETE PACKAGES | P |
| abortBroadcast | API | MODIFY PHONE STATE | P |
| pm install | CR | Ljavax crypto Cipher | API |
| READ SMS | P | WRITE CONTACTS | P |
| WRITE SMS | P | BIND INPUT METHOD | P |
| mount | CR | PROCESS OUTGOING CALLS | P |
| FACTORY TEST | P | SET WALLPAPER HINTS | P |
| WRITE APN SETTINGS | P | READ LOGS | P |
| RESTART PACKAGES | P | CALL PHONE | P |
| CHANGE COMPONENT ENABLED STATE | P | INTERNAL SYSTEM WINDOW | P |
| getSubscriberId | API | BLUETOOTH ADMIN | P |
| BIND REMOTEVIEWS | P | CHANGE WIFI MULTICAST STATE | P |
| DISABLE KEYGUARD | P | UPDATE DEVICE STATS | P |
| CHANGE WIFI STATE | P | RECEIVE BOOT COMPLETED | P |
| CLEAR APP CACHE | P | SecretKey | API |
| READ PHONE STATE | P | getLine1Number | API |
| TelephonyManager | API | BLUETOOTH | P |
| FindClass | API | DEVICE POWER | P |
| AUTHENTICATE ACCOUNTS | P | READ EXTERNAL STORAGE | P |
| chmod | CR | BROADCAST WAP PUSH | P |
| BIND WALLPAPER | P | FLASHLIGHT | P |
| BIND ACCESSIBILITY SERVICE | P | HARDWARE TEST | P |
| DELETE CACHE FILES | P | WRITE SECURE SETTINGS | P |
| GET PACKAGE SIZE | P | Runtime | API |
| READ CALL LOG | P | INTERNET | P |
| INSTALL PACKAGES | P | READ CONTACTS | P |
| GET ACCOUNTS | P | RECORD AUDIO | P |
| SMSReceiver | API | Intent.action.RUN | intent |
| Ljava net URLDecoder | API | REBOOT | P |
| intent.action.BOOT COMPLETED | Intent | ACCESS LOCATION EXTRA CS | P |
| GLOBAL SEARCH | P | READ HISTORY BOOKMARKS | P |
| MANAGE ACCOUNTS | P | getNetworkOperator | API |
| ACCESS NETWORK STATE | P | EXPAND STATUS BAR | P |
| SET ORIENTATION | P | jar | CR |
| /system/bin | CR | DexClassLoader | API |
| USE CREDENTIALS | P | WRITE HISTORY BOOKMARKS | P |
| RECEIVE WAP PUSH | P | CHANGE NETWORK STATE | P |
| bindService | API | getDeviceId | API |
| NFC | P | STATUS BAR | P |
| RECEIVE MMS | P | SET WALLPAPER | P |
| BIND APPWIDGET | P | HttpGet init | API |
| Ljavax crypto spec SecretKeySpec | API | getPackageManager | API |
| exec | API | getCallState | API |
| getSimSerialNumber | API | apk | CR |

P: permission

CR: Command related

API: API call related